\documentclass{elsarticle}

\usepackage[boxed]{algorithm2e}
\usepackage{amssymb}

\journal{Els}

\begin{document}
\begin{frontmatter}



\title{Variable Neighborhood Search for solving Bandwidth Coloring Problem}

\author[a1]{Dragan Mati\'c\corref{cor1}}
 \cortext[cor1]{Corresponding author.
phone/fax:+38751319142 }
\ead{matic.dragan@gmail.com}
\author[a2]{Jozef Kratica}
\ead{jkratica@gmail.com}
\author[a3]{Vladimir Filipovi\'c}

\address[a1]{University of Banjaluka, Faculty of Mathematics and Natural Sciences, Mladena Stojanovi\'ca 2, 78000 Banjaluka,
Bosnia and Herzegovina}
\address[a2]{Mathematical Institute of the Serbian Academy of Sciences and Arts, Belgrade, Serbia}
\address[a3]{University of Belgrade, Faculty of Mathematics, Studentski trg 16, 11000 Belgrade,
Serbia}

\begin{abstract}
This paper presents a variable neighborhood search (VNS) algorithm for solving bandwidth coloring problem (BCP) and  bandwidth multicoloring problem
(BMCP). BCP and BMCP are  generalizations of the well known vertex coloring problem and they are of a great interest  from both theoretical and
practical
points of view. Presented VNS combines a shaking procedure which  perturbs the colors for an increasing number of vertices and a specific variable
neighborhood descent (VND) procedure, based on the specially designed arrangement of the vertices which  are the subject of re-coloring. By this
approach,  local search  is split in a series of disjoint procedures, enabling better choice of the vertices which are addressed to re-color.
The experiments show that proposed method is highly competitive with the state-of-the-art algorithms and improves 2 out of 33 previous best known solutions for BMCP.
\end{abstract}
\end{frontmatter}

\section{Introduction}
\label{}
Vertex coloring problem (VCP) and its generalizations belong to a well known and widely researched class of graph coloring problems and therefore
represents
 a major challenge for most researchers in the field of combinatorial optimization. Various generalizations and variants of the VCP have been researched
 over the years and there are thousands of scientific papers proposing various  methods for solving VCP and its generalizations.

 In VCP, one needs to color the vertices of the graph in such a way that adjacent vertices must be colored with different colors and the aim is to minimize
 the number of used colors. During the years, as one of the most studied NP-hard combinatorial optimization problems, VCP has undergone many
 generalizations. This paper deals with two generalizations of the VCP: bandwidth coloring problem (BCP) and bandwidth multicoloring problem (BMCP).

 BCP is a straightforward generalization of the VCP, where for each two adjacent vertices $u$ and $v$ the distance, $d(u,v)$ is imposed and the difference
 between two colors assigned to $u$ and $v$ must be larger  than or equal to the defined distance.  The task is to color the vertices with the smallest
 number of colors.
 Formally, for a graph $G = (V, E)$ with positive integer distance function $d(u, v),\ (u,v) \in E$, the objective of the BCP is to find the coloring $c$
 such that for each pair of adjacent vertices $u$ and $v$, $|c(u)-c(v)|\geq d(u,v)$  and the number of colors is minimized. Obviously, if the distance
 between any pair of adjacent vertices is equal to 1, the BCP is brought down to the VCP.

 BMCP generalizes BCP by including the multicoloring of the vertices. For each vertex $v$ in the input graph a positive integer weight $w(v)$ is
 introduced, giving the requirement how many colors must be assigned to that vertex. Additionally, distance between the vertex to itself is also given,
 holding the condition inherited from BCP. Formally, for a given graph $G = (V, E)$ with positive integer distance function $d(u, v),\ (u,v) \in E$, with
 the condition that $u$ can be equal to $v$ (loops are allowed) and for the given positive integer value $w(u)$, for each vertex $u\in V$, the aim of the
 BMCP is to find the coloring $c$ such that each vertex $u$ is colored with  $w(u)$ colors, for each pair of adjacent vertices $u$ and $v$,
 $|c(u)-c(v)|\geq d(u,v)$ (including loops) and the number of colors is minimized. In a case when $w(u)=1$ for each vertex $u$, BMCP is reduced to the
 BCP.

BMCP can be converted into the BCP, by splitting each vertex $v$  into a clique of cardinality $w(v)$. Each edge in the clique is assigned the distance
$d(u,u)$, corresponding to the distance of the loop edge of the vertex $u$ in the original graph. By this approach, each instance of BMCP, with $n$ vertices, is transformed to
the instance of the BCP, having $\sum_{i=1}^n w(i)$ vertices.
  This fact leads to the approach of constructing the algorithm for solving  only the BCP which, after the  explicit or implicit
  construction of the appropriate graph,  can also be applied to solve BMCP.  Both BCP and BMCP are NP-hard problems, since they generalize VCP, which is
  NP-hard in a general case.

Like many other graph - based problems, VCP and its generalizations enjoys many practical applications.

For example, it is well known that timetabling problems can be interpreted as graph coloring problems.
Timetabling problems typically  includes the task of assigning timeslots to the events. In university timetabling problems, events (lectures or exams) are
interpreted as vertices, constraints by edges and timeslots by colors.
 According to \cite{Sch99}, whole class of  university timetabling problems can be roughly divided into the subclasses:  course timetabling problems (CTP)
 and examination timetabling problems (ETP).
For example, in CTP, lessons  are vertices and two vertices are adjacent if corresponding lessons are taught by the same professor. The task is to assign
the timeslots to the lessons (i.e. assign the colors to the vertices), in such a way that two lessons given by the same professor must not be taught at the
same time. In ETP, exams are interpreted by the vertices and two vertices are adjacent if there is at least one student passing the both exams. If a new
constraint is introduced by including the required time distance between two exams, the coloring problem which arises from this approach becomes BCP.
Additional constraints can be placed on the each vertex, for example, if one exam has to be organized for many groups of students. In that case, the
corresponding coloring problem is the BMCP.

Recent approaches  for solving timetabling problems, based on using graph coloring techniques, involves the hybridization of the methods automated by the
hyper-heuristic methods: a generic hyper-heuristic approach of various constructive heuristics for solving the VCP is presented in \cite{bur07}. In a
consecutive work \cite{qu09}, the authors presented a random iterative graph based hyper-heuristic, which adaptively hybridise  two graph colouring
heuristics at different stages of solution. Another recent hyper-heuristic approach \cite{sab12} utilizes the hierarchical hybridizations of four low level
graph coloring heuristics: largest degree, saturation degree, largest colored degree and largest enrollment. A constructive heuristic for finding a
feasible timetable is recently presented in \cite{bud12}. Much more information and a recent general survey of graph coloring algorithms can be found in
\cite{lew12}.

An intensive grow of the usage of the wireless communication protocols, like in mobile telephony, satellite communication, wireless sensor networks, radio
and TV broadcasting etc. leads to the development of various strategies applied to organize and manage the frequencies of the signals: the handling of
interference among radio signals, determining the availability of frequencies, as well as including various optimization criteria for achieving better
performances of the overall system. The optimization problems which arises from the task of managing frequencies  fall in the class of so called frequency
assignment problems (FAP) and according to the special needs of a concrete problem, many variants of the FAP appears. A common feature of most of the
variants includes the distance constraint imposed on pairs of frequencies, in order to avoid or reduce the interference between close communication
devices. In other words, two communication points which are close enough (as so can interference each other) must be assigned with the enough different
frequencies. Without any additional constraint, this frequency assignment problem is simply called feasible frequency assignment problem (F-FAP) and it is
obvious that this F-FAP corresponds to the finding of feasible coloring of the corresponding graph.

In connection with FAP problems, Hale \cite{hal80} introduced so called  T-coloring of the graph. For a given graph $G= (V,E)$ and for each edge
$\{u,v\}\in E$, we associate a set $T_{u,v}$ of nonnegative integers, containing 0.  A T-coloring of $G$ is a function (an assignment) $c:V(G)\rightarrow
\textbf{N}$ of colors to each vertex of $G$, so that if $u$ and $v$ are adjacent in $G$, then $|c(u) - c(v)|$ is not in $T_{i,j}$. In simple words, the
distance between two colors of adjacent vertices must not belong to the associated set  T. The span of a T-coloring $c$ is the difference between the
smallest and the highest color in $c$. The optimization version of T-coloring is finding the minimum span of all possible T - colorings. It is obvious  if
$T_{u,v}=\{0\}$ for each edge $\{u,v\}$, then problem becomes VCP.
 A generalization of T-coloring is a set T-coloring problem \cite{tes93}, which introduces the multiple coloring of vertices in the graph. For each vertex
 $v$, the set T-coloring problem includes the assignment of a nonnegative integer $\delta_v$, providing the information how many colors need to be assigned
 to $v$. Also, for each vertex $v$ a set $T_{v,v}$ is introduced. The constraints for each pair of adjacent vertices, previously introduced in T-coloring
 problem are extended with the constraints related to each vertex $v$: the distance of any two numbers (colors) assigned to   $v$  must not be in
 $T_{v,v}$.
BCP is a restriction of T-coloring, where the constraint on adjacent vertices is replaced by the proper condition of BCP: $|c(u)-c(v)|\geq t(u,v)$,
where $t(u,v)$ is a numerical value. Similarly, BMCP can be considered as an instance of the set T-coloring problem, where the set $T_{v,v}$ is replaced by
the numerical value, corresponding to $d(v,v)$. A survey of the results and  problems concerning with T-colorings can be found in \cite{rob91}.

According to the other special needs of a concrete problem, various subproblems of FAP can arise  \cite{aar07}:
\begin{itemize}
\item The Maximum Service (Max-FAP) and Minimum Blocking Frequency Assignment Problems  (MB-FAP) : if feasible solutions to the F-FAP are not available
    or are difficult to find, a partial solution that assigns as many frequencies as possible to the vertices can be considered. This problem is similar
    with registry allocation problem.
\item The Minimum Order FAP (MO-FAP): Frequencies are assigned in such a way that no unacceptable interference occurs, and the number of different used
    frequencies is minimized. This problem is a direct generalization of the graph coloring problem.
\item The Minimum Interference Frequency Assignment Problem (MI-FAP):    frequencies from a limited number of available frequencies are assigned in such
    a way that the total sum of weighted interference is minimized; in fact in this problem, we minimize the sum of the penalties incurred by the
    frequency choices
\item The Minimum Span Frequency Assignment Problem (MS-FAP): the problem is to assign frequencies in such a way that the interference between the points is
    avoided, and the difference between the maximum and minimum used frequency, the span, is minimized. This problem is equivalent to the BCP. In a case
    when the feature that one point is to be assigned with more than one frequency, the problem becomes the BMCP.
\item Other variants with some additional constraints

\end{itemize}

More applications, as well as other discussions considering graph coloring and its generalizations is out of the paper's scope and can be found for example
in \cite{lew12,mal10,gal13}.

The rest of this paper is organized as follows. The next section recapitulates previous work regarding BCP and BMCP. VNS approach is described in details
in the section \ref{VND}.
Section \ref{section:experimental} contains experimental results obtained on the instances from the literature, while the last section contains conclusions
and future work.

\section{Previous work on solving BCP and BMCP}

As stated above, BCP and BMCP, as well as MS-FAP has been intensively solved by a large number of successful methods. In 2002, in order to encourage research on computational methods for solving graph coloring problems, a "Computational Symposium on graph coloring and its generalization" was organized. The Symposium included the following topics: exact algorithms and heuristic approaches for solving the graph coloring problems, applications and instance generation, as well as methods for algorithm comparison \cite{tri02}. In order to make a fair comparison of the proposed methods, for testing the solving  the bandwidth (edge weights) multicoloring (node weights) or both graph coloring problems, a standard set of test instances was suggested. This set is also used in this paper.
At the Symposium, several successful heuristic methods were proposed. A problem-independent heuristic implementation called Discropt, designed for "black box optimization", was adapted  to graph coloring problems and provided a good test of the flexibility of the system \cite{pha02}. At the same symposium, Prestwitch (\cite{pre02}) proposed a heuristic based on the hybridization of local search and constraint propagation. In the consecutive contribution \cite{pre08}, the author extended his previous work by adding the constraint programming technique of forward checking in order to prune the colouration neighbourhoods. Hybrid methods using a squeaky-wheel optimization, combined with hill-climbing and
with tabu heuristic, were described in \cite{lim03} and \cite{lim05}.
 Chiarandini et al. in \cite{chi07} presented an experimental study of local search algorithms for solving general and large size instances of the set T -colouring problem.

 Malaguti and Toth (\cite{mal08}) proposed a successful method for solving BCP and BMCP, which combines an evolutionary algorithm with a tabu search procedure.  Like the method used in \cite{lim05}, Malaguti and Toth's  algorithm starts with the construction of an initial solution with a greedy approach. After that, it tries to improve the starting solution by reducing by one unit the number of colors used.
  Marti et al. \cite{mar10} proposed the memory-based and memory-less methods to solve the bandwidth colouring problem, based on tabu search and GRASP.
Paper presented by Lai and L\"u \cite{lai13} uses Multistart iterated tabu search (MITS) algorithm, which integrates an iterated tabu search with multistart method and a problem specific operator designed for the perturbation. This method uses previous known best results as a starting result and tries to improve it by decreasing it by one, repeating this process until no legal coloring can be found. The tabu search proposed in \cite{lai13} is successfully combined with a path relinking algorithm (paper available at ArXiv \cite{lai14}). Computational results demonstrate that the proposed algorithm outperforms previous methods, improving 15 out of 66 instances and matches 47 previous best known results. Lastly, a very recent work presented by Jin and Hao \cite{jin14} uses a learning-based hybrid search for solving BCP and BMCP. The algorithm firstly builds feasible partial colorings, which are further improved in a local search phase. In the construction phase, a learning-based guiding is used to determine the next vertex for color assignment, while in the local search phase tabu search technique is used to repair the solution. This approach further improves the best solutions for 14 common used instances.

From the last paragraph it is evident that in recent years the BCP and BMCP have been intensively solved by many highly efficient heuristics. As a consequence, the solutions for the benchmark data got by these methods are of a very high quality. In the next section, we describe the variable neighborhood search, which can further improve two of these intensively solved benchmark instances.


\section{VNS for solving BCP and BMCP}\label{VND}
This section presents the VNS for solving BCP and BMCP. Recall that each BMCP instance is implicitly transformed to an instance of BCP, by replacing each
vertex $v$ of the weight $w(v)$, by the clique of the size $w(v)$. Therefore, only the algorithm for solving BCP is described, but it is also applied on
the BMCP, after the mentioned transformation of the instances.

Variable Neighborhood Search (VNS) algorithm was originally described by Mladenovi\'c and Hansen (\cite{mla97,han05}).
In recent years, VNS has been proven as a very effective and adoptable metaheuristic, used for solving a wide range of complex optimization problems.
 The basic strategy of the VNS is to focus the investigation of the solutions which belong to some neighborhood of the current best one.
In order to avoid being trapped in local suboptimal solutions, VNS  changes the neighborhoods, directing the search in the promising and unexplored areas.
By this systematic change of  neighborhoods, VNS iteratively examines a sequence of neighbors of the current best solution, following the approach that
multiple local optima are often in a kind of correlation, holding the 'good parts' of the current best solution and trying to improve the rest of it.

Many successful implementations of the standard VNS, as well as its many variants, prove that this successive investigation of the quality of the current
solution's neighbors  can lead to better overall solutions. Standard VNS usually imposes two main procedures: shaking and local search (LS). Shaking
procedure manages the overall system of the neighborhoods and in each iteration suggests a new point (potential solution) from the current neighborhood. In
order to better widen the search, shaking procedure often uses the neighborhoods of the different cardinality.  More precisely, for the given numbers
$k_{min}$ and $k_{max}$, a system of neighborhoods $N_{k_{min}},N_{k_{min}+1},...,N_{k_{max}}$ is constructed. For each value $k\in [k_{min},k_{max}]$, a
(usually random) solution from the neighborhood $N_k$ is chosen, which is the subject of further possible improvement inside the LS procedure.

LS is trying to improve the suggested solution, by investigating the other solutions in its neighborhood, usually formed by some minor changes of it. Local
search is usually implemented by using either best improving strategy or first improving strategy. While in the best improving strategy local search
investigates all neighbors, keeping the current best one as a new solution, the first improving strategy stops when the first improving neighbor is found.

In the proposed VNS, we enhance the basic LS approach by "splitting" the local search in a disjoint union of smaller ones, enabling the algorithm to
"switch" between different neighborhoods inside the LS procedure. The current  neighborhood is analysed until the solution is improved. If so, it is
restarting from the first neighborhood, otherwise, goes into the next one. This approach requires the introduction of a specific variable neighborhood descent (VND) procedure which
manages the examination of the neighborhoods, allowing the algorithm to firstly analyse the best suitable one. The description of the VND is in details
explained in the section \ref{subsection:ls}.

The optimization process of the VNS algorithm finishes when the stopping criterion is achieved, usually given by the maximum number of iterations or
maximum time allowed for the execution and the latter is the case in this paper.

Some recent successful implementation of the VNS using the VND approach can be found for example in \cite{hu06,her01,her09}.

The overall VNS algorithm is shown on the figure \ref{fig:overall}. The algorithm inputs the following data: distance constraint matrix, values $k_{min}$ and $k_{max}$, denoting the minimal and maximal neighborhood structures, $time_{max}$ - maximal allowed execution time and the value $p_{move}$, representing the probability of shifting from one solution to another, in a case of equal objective functions. After the data input, VNS starts with a greedy heuristic (described in the subsection
\ref{subsection:greedy}), which gives  the upper bound ($UB$) for the total number of colors. After the $UB$  is determined, the initial value of $k^*$ is set to $UB$ (legal coloring with $UB$ colors), and the starting solution of the
VNS is constructed by the procedure Init(). Initial solution uses one color less than the value $UB$. This procedure, together with the objective function,
is described in the subsection \ref{subsection:init}. The minimization process is performed in the shaking procedure (subsection
\ref{subsection:shaking}) and the VND procedure, which is, together with the Compare procedure described in the subsection {\ref{subsection:ls}. During the minimization process inside the VND procedure the algorithm is trying to improve the solution to the feasible one. If this situation happens, the value $k^*$ is decreased by one, that legal coloring is remembered and the algorithm repeats the search process by decreasing the total number of colors.  The algorithm stops when  the
maximum execution time is reached. The result of the algorithm is the value $k^*$, i.e. the $k^*$-coloring of the given graph.
\begin{figure}

\begin{algorithm}[H]
    \KwData{Graph, Distance constraint matrix, $k_{min}$, $k_{max}$, $time_{max}$, $p_{move}$}

  \KwResult{number $k^*$}

   $xs\leftarrow$ GreedyHeuristic()\;
   $x\leftarrow $RemoveLastColor($xs$)\;
   $k \leftarrow k_{min}$\;
  \While{$time<time_{max}$}{
    $x'=$Shaking($x,k$)\;
    $x''$=VND($x$,$x'$)\;
    \eIf{Compare($x'',x,p_{move}$)}{$x\leftarrow x''$\;}
    {
      \eIf{$k<k_{max}$}{
      $k\leftarrow k+1$\;
      }
      {$k\leftarrow k_{min}$\;
      }
    }
    }

\end{algorithm} \caption{VNS pseudocode}\label{fig:overall}
\end{figure}

\subsection{Constructive heuristics}\label{subsection:greedy}
Before the minimization process is started,  an initial solution (legal $k$-coloring, for some $k$) needs to be constructed. Our algorithm uses the
greedy approach similar to the greedy algorithm used and minutely described in \cite{lim05,fij12}. This greedy algorithm  takes a sequence of 'split nodes',
greedily assigning colors to them. For each node a set of 'forbidden' colors is firstly formed  and after that the algorithm chooses the smallest color not belonging to that set.

In the literature, other approaches for getting starting solutions  are also used. For example, Malaguti and Toth in \cite{mal08}, considered several
greedy algorithms proposed for solving VCP: sequential greedy algorithm (SEQ), as well as another greedy approach - DSATUR from \cite{bre77}, similar to
SEQ, but one which dynamically chooses the vertex to color next, i.e. the one which minimizes a given score.  In order to fast compute an initial upper
bound, Malaguti and Toth performed 20 runs of the greedy algorithm SEQ.

 Malaguti and Toth's greedy approach in most case instances achieves better (lower) upper bounds  than the greedy algorithm we took from  \cite{fij12}, but the
 experiments indicate that presented VNS easily decrease those (higher) upper bounds. Therefore, using  slightly greater starting values of $k$ could not
 significantly aggravate the overall optimization process. The exceptions of this "rule" can appear in some cases of small instances, where Malaguti and Toth's
 greedy algorithm achieves nearly best known solutions, so iteration process needs to decrease the upper bound only for few values. In
 these cases, our algorithm needs some more time, since it starts with higher upper bounds.

In the most recent papers (\cite{lai13,lai14,jin14}), a constructive heuristic is not used. The proposed algorithms simply start with the best known values of $k$ from the literature as the starting values and try to construct
the feasible $k$-coloring. If succeeds, i.e. the legal $k$-coloring is found, the algorithms decrease the value $k$ by 1 and try to find the $k-1$ coloring. This iterative
process stops when no legal coloring can be found. Although this approach appears to be very successful and can speed
up the overall process (because the algorithms do not spend any time to construct some starting solution and decrease it many times in the optimization
process), we still decided to follow the approach used in \cite{mal08} and  \cite{fij12}: our basic approach is to construct an initial solution by greedy algorithm and decrease it in
the iteration process, while the  stopping criteria are not satisfied.

\subsection{The initialization and the objective function}\label{subsection:init}
For the given graph $G=(V,E)$, the solution is represented by an integer array of the dimension $n,\ n=|V|$.
After the upper bound ($UB$) is determined by the greedy approach, the VNS starts with the constructing the initial solution.
 Each vertex is randomly colored by a color chosen from the interval $[1,UB-1]$. From this representation, it is obvious that the algorithm deals both with
 feasible and infeasible solutions. So, in the objective function, each solution is a subject of the evaluation, where the objective value is proportional
 to the "degree of infeasibility". In order to construct the appropriate function, we took into account not only the number of conflicts (pairs of adjacent
 vertices $i$ and $j$ for which a conflict appears) but also the "degree" of each conflict, i.e. the difference between the given distance and the distance
 between two assigned colors. Similar approach can also be seen for example in \cite{lai13,fij12}.

For the given solution $x$, represented by the array $[c(1),c(2),...,c(n)]$, where $c(i),\ i=1..n$ is the color assigned to the vertex $i$, the objective function Calculate() is defined as follows:
\begin{equation}\label{eqn:obj}
Calculate(x) = \sum_{\{i,j\}\in E} \max \{{0, d(i,j)-|c(i)-c(j)|}\}
\end{equation}
where $d(i,j)$ is the given distance between the vertices $i$ and $j$.
From the equation \ref{eqn:obj} it can be seen that  the infeasible solutions are penalized by increasing the value of the objective function, if the distance
conflicts appear. In a case when there are no distance conflicts, the objective function is equal to 0 and in that case the solution is feasible i.e. the
legal coloring is found.
Additionally to the calculating the objective function, in the procedure Calculate() for each vertex $v$ we remember the value $conflicts(v)$ containing
the value of distance conflicts of $v$. More precisely, this value is calculated by the formula:
\begin{equation}\label{eqn:conflits}
conflicts(v) = \sum_{j:\{v,j\}\in E} \max \{{0, d(v,j)-|c(v)-c(j)|}\}
\end{equation}

These values, obtained for each vertex $v\in V$, take places  for the arrangement of the vertices, performed inside the VND procedure.
\subsection{Neighborhoods and Shaking procedure} \label{subsection:shaking}
The shaking procedure creates a
new solution $x'$, ($x' \in N_k(x)$), which is based on the current best
solution $x$.

In order to define the $k$-th neighborhood we use the following procedure. Some
$k$ vertices  from $V$ are chosen randomly and for each chosen
vertex, the color is randomly changed to some other color from the interval $[1,max\_color]$, where $max\_color$ is the maximal color used in the coloring. The solution $x'$, formed in the described way is the subject of the further improvements in the VND.

In the algorithm, the value $k_{min}$ is set to 2. Experiments show that the algorithm achieves best performances for  $k_{max}=20$.

\subsection{Variable neighborhood descent} \label{subsection:ls}
After the solution $x'$ is obtained by the shaking procedure, a series of special designed local
search procedures  is called.
In each call, the ordering of vertices, which are the subject of change in the VND procedure, is determined. This ordering enables the vertices with more
conflicts to earlier become the subject of improvement. This number of conflicts is calculated for each vertex each time when the function Calculate() is
called (see the subsection \ref{subsection:init}). This strategy avoids the unnecessary steps in the local search, by increasing the probability that the
solution will be earlier improved  if the "worse" vertices are considered first, rather than if the vertices are considered sequentially without ordering. At
the other hand, experiments show that this criterion is not enough for getting a quality ordering, since many vertices, after a few iterations, have the
same number of conflicts (in the sense of the value $conflicts(v)$). So, additional criteria are involved. Firstly, we started from the fact that the
colors used in BCP are not equal among themselves (like in the case of VCP). In fact, the colors which are closer to the middle of the interval $[1,nc]$
($nc$ is total number of used colors) are harder to be replaced by another, comparing to the colors far off-the middle. So, if two vertices $u$ and $v$
have the same number of conflicts ($conflicts(u) = conflicts(v)$), we firstly choose $u$ if $|nc/2-c(u)|<|nc/2-c(v)|$, otherwise we firstly choose $v$.

In order to further improve the behaviour of the VND, we involve one additional criterion for the arrangement of the vertices, in cases when first two
criteria do not make any difference between vertices. The third criterion is based on two components: for each vertex $v$ we calculate the value
$weights(v)$, as the  sum of the weights (distances) of the edges incident with $v$. Additionally, for the vertex $v$, we take into account the maximal
edge weight (distance) for the vertex $v$ ($maxw(v)$). Finally, the third criterion is calculated by the geometric mean of these two values, i.e. in cases
when first two criteria do not determine the priority,  the vertex $u$ is chosen before the vertex $v$ if $\sqrt{weights(u)\cdot
maxw(u)}>\sqrt{weights(v)\cdot maxw(v)}$. Otherwise, the VND firstly choose $v$.
Although the  third criterion  influence on the ordering relatively rarely, the experiments show that it refines the  ordering of vertices in a good
direction and improves the overall VND.

 Beside the main experiments described in the sections \ref{section:bcpexperimental} and \ref{section:bmcpexperimental}, in order to  justify the usage of these three criteria, additional experiments are performed. In the experiments (described in the section \ref{section:criteriaexperimental}), various combinations of the mentioned criteria are considered. The obtained results indicate that the  approach of using all three criteria generally provides better results than variants in which some of these criteria are omitted.

The pseudocode of the VND procedure is shown in the figure \ref{fig:ls}.

\begin{figure}

\begin{algorithm}[H]
    \KwData{$x$,$x'$}
 $impr \leftarrow true$\;
 $x''\leftarrow x'$\;
  \While{$impr$}{
     $impr  \leftarrow false$\;
     $objval \leftarrow$  ObjF($x''$)\;
     $vertices \leftarrow$ qsort($vertices$,$criteria$)\;
     \ForEach{vertex $v\in vertices$}{
        uncolor($v$)\;
        remove\_conflicts($x''$,$v$)\;
         \ForEach{color $c$}{
            recolor($v$,$c$)\;
            calculate\_new\_conflicts($x''$)\;
           }
           $x_t''$ = find\_best\_recoloring($x''$)\;
           \eIf{ (ObjF($x_t''$)==0)}{ 
					      //feasible coloring is found\;
                $x''\leftarrow x_t''$\;
                $impr \leftarrow true$\;
                $xs\leftarrow x_t''$\;
                $k^* \leftarrow $max\_color($x_t''$)\;
                remove\_last\_color($x''$)\;
           }
					{
           \If{ (ObjF($x_t''$)$<objval$ )}{ //improvement happened\;
             $x''\leftarrow x_t''$\;
             $impr \leftarrow true$\;
             $objval$ = ObjF($x_t''$)\;
            }
     }
    }
}   
\end{algorithm} \caption{VND pseudocode}\label{fig:ls}
\end{figure}

After the objective function for the solution $x''$ is calculated by the function ObjF() and
the array of vertices is  arranged by using the mentioned criteria, the VND iteratively chooses the vertices from that array. For the selected vertex
$v$, the VND is trying to find "better" color in the following way. If the vertex $v$ is currently colored by the color $c(v)$, we first "uncolor" that
vertex, also removing the conflicts which arise by using that color for the vertex $v$. It should be noted that we do not need to calculate the objective
function from the beginning, since only conflicts related to the chosen vertex influence on the total sum. Therefore, after only these conflicts are
removed, we iteratively color the vertex $v$ by all other colors, trying to find better coloring. Simultaneously, we calculate the new conflicts, which
appear by new coloring and remember these values (conflicts). After we tried all the colors for the vertex $v$, we choose the one, which generates the
least total sum of conflicts. In that way, we get a new temporary solution $x_t''$ and three possibilities can happen:

\begin{itemize}
\item The objective value of $x_t''$  is equal to 0: That means that the VND was totally successful and not just improved the objective value, but gave
    the feasible coloring  with the less number of colors than the previous best one. In this case, we remember that  coloring, set $xs=x_t''$ and
    continue the search as follows: we set up a new current solution $x''=x_t''$ and replace the maximal color in $x''$ by some other, randomly chosen
    color. The VND, i.e. the improvement process is then applied again on the current $x''$ and  the next vertex. It should be noted that if this case
    happen,  the maximum number of colors may be decreased by more than one in one such step.
\item The objective value of $x_t''$ is greater than 0, but still less than the starting one: we hold the changes arisen in the current step (we set
    $x''=x_t''$) and continue the VND  with the next vertex.
\item The objective value of $x_t''$ is greater or equal than the starting one: we continue the VND with $x''$ and the next vertex, without any change.

\end{itemize}
The VND finishes when no more improvements can be done and the algorithm analysis the results of the VND in the procedure Compare:  In  cases when number of colors used in the
solution $x''$ is less than in the solution $x$, or if the objective value of $x''$ is less than of the $x$, the currently best solution $x$ gets the value
$x''$. If the objective values of the two solutions $x$ and $x''$ are the same, then   $x = x''$ is set with probability $p_{move}$ and the algorithm
continues the search with the same neighborhood.
In all other cases, the search is repeated with the same $x$ and the next neighborhood.

Although some VNS implementation allows low values of $p_{move}$, even $p_{move}=0$ (see for example \cite{mat12}), the experiments indicate that the
proposed algorithm for solving BCP achieves best performances for the value $p_{move} = 0.5$. This  can be explained by the integer nature of the objective
function and the fact that there is a huge number of solutions with the same objective value (especially when the objective value is decreased to 1). In
general, higher values of $p_{move}$  increase the probability of the appearance of cycles. For the value $p_{move} = 0.5$, the probability of the
appearance of the cycle of the length $l$ is equal to $0.5^l$ and it is rather low, even for the smaller values of $l$.
\section{Experimental results}\label{section:experimental}

This section presents the experimental results, which show the
effectiveness of the proposed VNS.
All the tests are carried out on the Intel i7-4770 CPU @3.40 GHz
with 8 GB RAM and Windows 7 operating system. For each execution, only one thread/processor is used. The VNS is implemented in C programming language and
compiled with Visual Studio 2010 compiler.

For all experiments we used standard set of GEOM instances, which consists of 33 geometric graphs generated by Michael Trick, available in \cite{tri02}.
In each GEOM instance, points are generated in a 10,000 by 10,000 grid and are connected by an edge if they are close enough together.
Edge weights are inversely proportional to the distance between vertices, which simulates the real situations where closer adjacent vertices
stronger interference each other. This set contains three types of graphs: for each dimension, one sparse (GEOMn) and two denser graphs (GEOMna and GEOMnb)
are given. The instances are originally generated for BCP, but they are also transformed for solving BMCP,  by introducing  the weights of vertices.  For
BMCP, vertex weights are uniformly randomly generated, between 1 and 10 for sets GEOMn and GEOMna, and between 1 and 3 for set GEOMnb.

Since there are several state-of-the-art algorithms in the literature, we could not establish a unique set of algorithm parameters to make a complete fair comparison to all these methods.  Anyway, we decided to follow similar  conditions from the most recent and most
successful approaches \cite{mal08,lai13,lai14,jin14}.  To check the speed of the
our computer CPU, we used a standard benchmark program (dfmax), together with a benchmark instance (R500.5) also used in the reference works. For this instance, we
report the computing time of 8 seconds, which is similar to the times reported in other recent works. Similar to the approach in \cite{lai14}, we set the timeout limits to 2 hours for  BCP instances and 4 hours for MBCP instances.  The algorithm stops if it cannot solve the coloring within the time limit. For each instance, we performed 30 independent executions, which is also the case in \cite{lai13}.

\subsection{Experimental results on BCP instances}\label{section:bcpexperimental}

This section reports the experimental results obtained on the first set of instances, related to the BCP problem. Table \ref{table:BCP1} provide the
results obtained by  the presented VNS. The data in the table are organized as follows: first three columns contain the instance name, number of vertices
($|V|$) and number of edges ($|E|$). The fourth column contain the best known result from the literature. The next five columns contain data related to the
VNS: column ($k^*$) contains the best found result, the average result (column $avg$) obtained in 30 runs, the total average execution time in seconds
needed to achieve the presented best result (column $time$), the hit rate ($N_{hit}$), as well as the column ($k^*-best$), which contains the difference
between  the best result obtained by the VNS and the previous best-known result from the literature.  A new best result is marked in bold.
\begin{table}\caption{\small{Results of the VNS obtained on BCP instances}}

\tiny\centering

\begin{tabular}{rrrrrrrrr}\hline
Instance & $|V|$ & $|E|$ & best & $k^*$ & avg        & time{[}s{]} & $N_{hit}$ & $k^*-best$ \\\hline
GEOM20   & 20    & 40    & 20   & 21    & 21         & 0.00643     & 30/30     & 1                 \\
GEOM20a  & 20    & 57    & 20   & 20    & 20         & 0.01507     & 30/30     & 0                 \\
GEOM20b  & 20    & 52    & 13   & 13    & 13         & 0.0053      & 30/30     & 0                 \\
GEOM30   & 30    & 80    & 27   & 28    & 28         & 0.01023     & 30/30     & 1                 \\
GEOM30a  & 30    & 111   & 27   & 27    & 27         & 0.086       & 30/30     & 0                 \\
GEOM30b  & 30    & 111   & 26   & 26    & 26         & 0.00817     & 30/30     & 0                 \\
GEOM40   & 40    & 118   & 27   & 28    & 28         & 0.01657     & 30/30     & 1                 \\
GEOM40a  & 40    & 186   & 37   & 37    & 37         & 1.21        & 30/30     & 0                 \\
GEOM40b  & 40    & 197   & 33   & 33    & 33         & 2.9218      & 30/30     & 0                 \\
GEOM50   & 50    & 177   & 28   & 28    & 28         & 0.03407     & 30/30     & 0                 \\
GEOM50a  & 50    & 288   & 50   & 50    & 50         & 4.2695      & 30/30     & 0                 \\
GEOM50b  & 50    & 299   & 35   & 35    & 35.0333333 & 716.35247   & 29/30     & 0                 \\
GEOM60   & 60    & 245   & 33   & 33    & 33         & 0.15027     & 30/30     & 0                 \\
GEOM60a  & 60    & 399   & 50   & 50    & 50         & 23.6351     & 30/30     & 0                 \\
GEOM60b  & 60    & 426   & 41   & 41    & 41.8       & 6430.65223  & 7/30      & 0                 \\
GEOM70   & 70    & 337   & 38   & 38    & 38         & 0.32483     & 30/30     & 0                 \\
GEOM70a  & 70    & 529   & 61   & 61    & 61.0666667 & 1513.0634   & 28/30     & 0                 \\
GEOM70b  & 70    & 558   & 47   & 48    & 49.2333333 & 7702.62073  & 2/30      & 1                 \\
GEOM80   & 80    & 429   & 41   & 41    & 41         & 2.59037     & 30/30     & 0                 \\
GEOM80a  & 80    & 692   & 63   & 63    & 63.3666667 & 3487.41707  & 21/30     & 0                 \\
GEOM80b  & 80    & 743   & 60   & 60    & 61.8666667 & 7851.79623  & 2/30      & 0                 \\
GEOM90   & 90    & 531   & 46   & 46    & 46         & 1.5592      & 30/30     & 0                 \\
GEOM90a  & 90    & 879   & 63   & 63    & 64.1333333 & 8082.11243  & 5/30      & 0                 \\
GEOM90b  & 90    & 950   & 69   & 71    & 72.7       & 7627.29127  & 4/30      & 2                 \\
GEOM100  & 100   & 647   & 50   & 50    & 50         & 320.8932    & 30/30     & 0                 \\
GEOM100a & 100   & 1092  & 67   & 68    & 69.5666667 & 8227.2824   & 4/30      & 1                 \\
GEOM100b & 100   & 1150  & 72   & 73    & 75.3666667 & 7945.8779   & 4/30      & 1                 \\
GEOM110  & 110   & 748   & 50   & 50    & 50.0333333 & 943.76663   & 29/30     & 0                 \\
GEOM110a & 110   & 1317  & 71   & 71    & 72.6333333 & 9152.86497  & 2/30      & 0                 \\
GEOM110b & 110   & 1366  & 77   & 79    & 81.0666667 & 8829.13587  & 2/30      & 2                 \\
GEOM120  & 120   & 893   & 59   & 59    & 59.0333333 & 615.18443   & 29/30     & 0                 \\
GEOM120a & 120   & 1554  & 82   & 82    & 84.2666667 & 9112.61543  & 2/30      & 0                 \\
GEOM120b & 120   & 1611  & 84   & 86    & 87.9333333 & 9159.90177  & 2/30      & 2 \\\hline
\end{tabular}

\label{table:BCP1}

\end{table} 
From the Table \ref{table:BCP1}, it can be seen  that  for the most of the smaller instances up to 60 vertices, the proposed VNS algorithm achieves best
known solutions, with the exception of three instances, for which in \cite{pha02} better results were reported. It should be mentioned that other
state-of-art algorithms could not achieve these results from  \cite{pha02} either. Since heuristic methods do not guarantee the optimality of the obtained
solutions, it could be interesting if an exact method appears to check these three solutions. For the rest of the instances (total of 18 middle and larger
instances), VNS achieves 12 best results and  in 6 out of 18 cases VNS
is not able to reach previously best-known solution from the literature.

The proposed VNS   solves the sparse graphs (GEOMn) easier than the dense ones (GEOMna and GEOMnb). From the column $N_{hit}$, it can bee seen that the
ratio between the number of runs of the VNS  when the previously best-known solution is reached and the total number of runs (30) is rather high for sparse
graphs.

Table  \ref{table:BCP2} describes the comparison of our approach to the state-of-the-art methods. The data are organized as follows: The first column
contain the instance name. The rest of the table contain data related to the approaches:
 the best result of the Discropt general heuristics (DGH) presented by Phan and Skiena \cite{pha02} (the execution time is not reported),
 Prestwich's  forward checking colouration neighborhood search (FCNS) from \cite{pre08}, Malaguti and Toth's
evolutionary algorithm (EA) from \cite{mal08}, the MITS algorithm presented by Lai and Lu from \cite{lai13}, path relinking (PR) algorithm from \cite{lai14} and learning hybrid-based search (LHS) from \cite{jin14}. All these results are extracted from \cite{jin14}. Last three columns
contain best results, the execution times for the proposed VNS and the difference between best VNS and previous best known result. It should be noted that the execution times are very different and are not reasonable comparable, since the algorithms achieve different $k$- colorings and the algorithms were run on the machines with different CPU speeds.

\begin{table}[htbp]
  \centering
  \caption{\small{Comparison of the proposed VNS algorithm to other reference works on BCP instances}}
  \tiny

\begin{tabular}{rrrrrrrrrrrrrrrl}\hline
         &  & DGH \cite{pha02}    & \multicolumn{2}{c}{FCNS \cite{pre08}  }& \multicolumn{2}{c}{EA \cite{mal08} }& \multicolumn{2}{c}{ MITS \cite{lai13}   }& \multicolumn{2}{c}{ PR \cite{lai14}   }
 & \multicolumn{2}{c}{ LHS \cite{jin14}   } & \multicolumn{2}{c}{ VNS   }&\\

Instance & best & $k$ & $k$ & time{[}s{]} &    & time{[}s{]} & $k$ & time{[}s{]} & $k$ & time{[}s{]} & $k$ & time{[}s{]} & $k^*$ & time{[}s{]} & diff \\
GEOM20   & 20   & 20  & 21  & 0           & 21 & 0           & -   & -           & 21  & 0           & 21  & 0           & 21    & 0.00643     & 1          \\
GEOM20a  & 20   & 20  & 20  & 0           & 20 & 0           & 20  & 0           & 20  & 0           & 20  & 0           & 20    & 0.01507     & 0          \\
GEOM20b  & 13   & 13  & 13  & 0           & 13 & 0           & 13  & 0           & 13  & 0           & 13  & 0           & 13    & 0.0053      & 0          \\
GEOM30   & 27   & 27  & 28  & 0           & 28 & 0           & -   & -           & 28  & 0           & 28  & 0           & 28    & 0.01023     & 1          \\
GEOM30a  & 27   & 27  & 27  & 0           & 27 & 0           & 27  & 0           & 27  & 0           & 27  & 0           & 27    & 0.086       & 0          \\
GEOM30b  & 26   & 26  & 26  & 0           & 26 & 0           & 26  & 0           & 26  & 0           & 26  & 0           & 26    & 0.00817     & 0          \\
GEOM40   & 27   & 27  & 28  & 0           & 28 & 0           & -   & -           & 28  & 0           & 28  & 0           & 28    & 0.01657     & 1          \\
GEOM40a  & 37   & 38  & 37  & 0           & 37 & 0           & 37  & 0           & 37  & 0           & 37  & 0           & 37    & 1.21        & 0          \\
GEOM40b  & 33   & 36  & 33  & 0           & 33 & 0           & 33  & 0           & 33  & 0           & 33  & 0           & 33    & 2.9218      & 0          \\
GEOM50   & 28   & 29  & 28  & 0           & 28 & 0           & 28  & 0           & 28  & 0           & 28  & 0           & 28    & 0.03407     & 0          \\
GEOM50a  & 50   & 54  & 50  & 2           & 50 & 0           & 50  & 0           & 50  & 0           & 50  & 0.1         & 50    & 4.2695      & 0          \\
GEOM50b  & 35   & 40  & 35  & 0           & 35 & 0           & 35  & 3           & 35  & 1           & 35  & 1.2         & 35    & 716.35247   & 0          \\
GEOM60   & 33   & 34  & 33  & 0           & 33 & 0           & 33  & 0           & 33  & 0           & 33  & 0           & 33    & 0.15027     & 0          \\
GEOM60a  & 50   & 54  & 50  & 1           & 50 & 0           & 50  & 1           & 50  & 0           & 50  & 0.1         & 50    & 23.6351     & 0          \\
GEOM60b  & 41   & 47  & 43  & 0           & 41 & 29          & 41  & 277         & 41  & 105         & 41  & 214.7       & 41    & 6430.65223  & 0          \\
GEOM70   & 38   & 40  & 38  & 0           & 38 & 0           & 38  & 0           & 38  & 0           & 38  & 0           & 38    & 0.32483     & 0          \\
GEOM70a  & 61   & 64  & 62  & 2           & 61 & 12          & 61  & 45          & 61  & 47          & 61  & 23.7        & 61    & 1513.0634   & 0          \\
GEOM70b  & 47   & 54  & 48  & 1           & 48 & 52          & 47  & 8685        & 47  & 6678        & 47  & 665.4       & 48    & 7702.62073  & 1          \\
GEOM80   & 41   & 44  & 41  & 0           & 41 & 0           & 41  & 0           & 41  & 0           & 41  & 0.1         & 41    & 2.59037     & 0          \\
GEOM80a  & 63   & 69  & 63  & 12          & 63 & 150         & 63  & 21          & 63  & 12          & 63  & 6.6         & 63    & 3487.41707  & 0          \\
GEOM80b  & 60   & 70  & 61  & 0           & 60 & 145         & 60  & 322         & 60  & 191         & 60  & 19.9        & 60    & 7851.79623  & 0          \\
GEOM90   & 46   & 48  & 46  & 3           & 46 & 0           & 46  & 0           & 46  & 0           & 46  & 0           & 46    & 1.5592      & 0          \\
GEOM90a  & 63   & 74  & 64  & 2           & 63 & 150         & 63  & 230         & 63  & 191         & 63  & 23.8        & 63    & 8082.11243  & 0          \\
GEOM90b  & 69   & 83  & 72  & 2           & 70 & 1031        & 69  & 20144       & 69  & 23560       & 69  & 779.2       & 71    & 7627.29127  & 2          \\
GEOM100  & 50   & 55  & 50  & 0           & 50 & 2           & 50  & 2           & 50  & 2           & 50  & 1           & 50    & 320.8932    & 0          \\
GEOM100a & 67   & 84  & 68  & 9           & 68 & 273         & 67  & 11407       & 67  & 5556        & 67  & 1557.4      & 68    & 8227.2824   & 1          \\
GEOM100b & 71   & 87  & 73  & 15          & 73 & 597         & 72  & 24561       & 72  & 41832       & 71  & 2038.6      & 73    & 7945.8779   & 2         \\
GEOM110  & 50   & 59  & 50  & 4           & 50 & 3           & 50  & 2           & 50  & 5           & 50  & 1.3         & 50    & 943.76663   & 0          \\
GEOM110a & 71   & 88  & 73  & 7           & 72 & 171         & 72  & 1529        & 71  & 5140        & 71  & 2218.7      & 71    & 9152.86497  & 0          \\
GEOM110b & 77   & 87  & 79  & 2           & 78 & 676         & 78  & 24416       & 78  & 18136       & 77  & 2598.7      & 79    & 8829.13587  & 2          \\
GEOM120  & 59   & 67  & 60  & 4           & 59 & 0           & 59  & 1           & 59  & 2           & 59  & 0.5         & 59    & 615.18443   & 0          \\
GEOM120a & 82   & 101 & 84  & 4           & 84 & 614         & 82  & 34176       & 82  & 62876       & 82  & 171.1       & 82    & 9112.61543  & 0          \\
GEOM120b & 84   & 103 & 86  & 9           & 84 & 857         & -   & -           & 85  & 66301       & 84  & 3568.7      & 86    & 9159.90177  & 2

\end{tabular}

\label{table:BCP2}

\end{table}

 From the Table \ref{table:BCP2}, it can be seen that the  VNS algorithm is  competitive with other  methods. VNS achieves total of 24 previous best-known solutions. It is evident that the results achieved by older methods from \cite{pha02} and \cite{pre08} are improved by the newer ones, with the exception
   of three small instances, for which the best results are reported in \cite{pha02}.
   The VNS and EA from \cite{mal08} give 28 same best results, in 3 cases EA is better and VNS is better
in two cases.  Comparing to the MITS from \cite{lai13}, for VNS and  MITS  23 equal best results
are reported, in 5 cases MITS is better, in one case VNS is better. In four cases, no solution is reported in \cite{lai13}. VNS achieve 27 same solutions as the two most recent methods PR and LHS, and in 6 cases PR and LHS are better than VNS.

\subsection{Experimental results on BMCP instances}\label{section:bmcpexperimental}
Algorithm developed for BCP instances can also be applied for solving BMCP, after the implicit transformation of each BMCP to BCP instance. From the
experimental results presented in this section we see that our VNS achieve many previously known  best known results and two new best ones.

Table \ref{table:BMCP1} provides the results obtained on BMCP instances. The Table \ref{table:BMCP1} is organized similar to the case of BCP instances:
first four columns contain the instance name, number of vertices ($|V|$) and number of edges ($|E|$) and the best known result from the literature. The
next five columns contain data related to the VNS: column ($k^*$) contains the best found result, the average result (column $avg$) obtained in 30 runs,
the total average execution time in seconds needed to achieve the presented best result (column $time$), the hit rate ($N_{hit}$), as well as the column
($k^*-best$), which contains the difference between  the best result obtained by the VNS and the  previous best-known result from the literature.  In
column $k^*$, new best results are marked in bold.

\begin{table}\caption{\small{Results of the  VNS obtained on BMCP instances}}
\tiny
\begin{tabular}{rrrrrrrrl}

Instance & $|V|$ & $|E|$ & best & $k^*$          & avg         & time{[}s{]} & $N_{hit}$ & $k^*-best$ \\\hline \\
GEOM20   & 20    & 40    & 149  & 149            & 149         & 54.2701     & 30/30     & 0                   \\
GEOM20a  & 20    & 57    & 169  & 169            & 169         & 3289.39977  & 30/30     & 0                   \\
GEOM20b  & 20    & 52    & 44   & 44             & 44          & 0.02153     & 30/30     & 0                   \\
GEOM30   & 30    & 80    & 160  & 160            & 160         & 5.7033      & 30/30     & 0                   \\
GEOM30a  & 30    & 111   & 209  & 209            & 209         & 4123.7125   & 30/30     & 0                   \\
GEOM30b  & 30    & 111   & 77   & 77             & 77          & 1.2923      & 30/30     & 0                   \\
GEOM40   & 40    & 118   & 167  & 167            & 167.0333333 & 2107.13567  & 29/30     & 0                   \\
GEOM40a  & 40    & 186   & 213  & 213            & 214.1666667 & 13192.2284  & 5/30      & 0                   \\
GEOM40b  & 40    & 197   & 74   & 74             & 74.0333333  & 1821.8568   & 29/30     & 0                   \\
GEOM50   & 50    & 177   & 224  & 224            & 224.1       & 1671.3315   & 27/30     & 0                   \\
GEOM50a  & 50    & 288   & 311  & 311            & 314.2333333 & 21685.93933 & 2/30      & 0                   \\
GEOM50b  & 50    & 299   & 83   & 83             & 84.2        & 14260.65607 & 5/30      & 0                   \\
GEOM60   & 60    & 245   & 258  & 258            & 258.4333333 & 5780.4842   & 19/30     & 0                   \\
GEOM60a  & 60    & 399   & 353  & 353            & 355.3       & 23989.30317 & 2/30      & 0                   \\
GEOM60b  & 60    & 426   & 113  & 114            & 115.7       & 16381.0242  & 2/30      & 1                   \\
GEOM70   & 70    & 337   & 266  & 267            & 267.7       & 12384.2772  & 14/30     & 1                   \\
GEOM70a  & 70    & 529   & 465  & \textbf{463} & 465.5333333 & 20686.36647 & 2/30      & -2                  \\
GEOM70b  & 70    & 558   & 115  & 116            & 118.2       & 16783.2271  & 1/30      & 1                   \\
GEOM80   & 80    & 429   & 379  & 379            & 381.0333333 & 16538.80857 & 2/30      & 0                   \\
GEOM80a  & 80    & 692   & 357  & \textbf{355} & 358.5333333 & 29208.93163 & 1/30      & -2                  \\
GEOM80b  & 80    & 743   & 138  & 138            & 139.0666667 & 13704.121   & 9/30      & 0                   \\
GEOM90   & 90    & 531   & 328  & 329            & 330.8333333 & 18760.85243 & 2/30      & 1                   \\
GEOM90a  & 90    & 879   & 372  & 373            & 374.9       & 24087.21087 & 2/30      & 1                   \\
GEOM90b  & 90    & 950   & 142  & 142            & 145.0333333 & 19996.89127 & 2/30      & 0                   \\
GEOM100  & 100   & 647   & 404  & 404            & 405.9333    & 14816.92453 & 6/30      & 0                   \\
GEOM100a & 100   & 1092  & 429  & 429            & 431.9333333 & 35663.42977 & 2/30      & 0                   \\
GEOM100b & 100   & 1150  & 153  & 156            & 159.3666667 & 24776.3847  & 1/30      & 3                   \\
GEOM110  & 110   & 748   & 375  & 375            & 376.2333333 & 22997.44417 & 8/30      & 0                   \\
GEOM110a & 110   & 1317  & 478  & 480            & 484         & 46954.99173 & 1/30      & 2                   \\
GEOM110b & 110   & 1366  & 201  & 202            & 203.4       & 22076.94307 & 4/30      & 1                   \\
GEOM120  & 120   & 893   & 396  & 396            & 398.5       & 17919.32823 & 8/30      & 0                   \\
GEOM120a & 120   & 1554  & 536  & 539            & 545         & 59717.3869  & 3/30      & 3                   \\
GEOM120b & 120   & 1611  & 187  & 190            & 192.5       & 22835.9246  & 3/30      & 3

\end{tabular}

\label{table:BMCP1}

\end{table} 
Data in Table \ref{table:BMCP1} indicate that VNS achieves 21 best known results and in 2 cases VNS obtains new best results. In 10 cases, VNS could  achieve
results close to the best ones.
For all smaller BMCP instances (up to 50 vertices), VNS obtains all previous best-known results, with a relatively high hit rate for most of them.
VNS achieves 2 new best results, for two middle instances GEOM70a and GEOM80a.  For 9 large-sized instances, (100-120 vertices), VNS achieves previously known best results in 4 cases and in 5 cases VNS achieves nearly best results.

Table  \ref{table:BMCP2} shows the comparative results obtained by the state-of-the-art methods from the literature and the proposed VNS.  The first column
of the table contains the instance name. The next five blocks of two columns contain best results and execution times of the five recent and most successful methods from the literature: FCNS \cite{pre08} by  Prestwich,
 Malaguti and Toth's EA from \cite{mal08}, the MITS
algorithm from \cite{lai13} presented by Lai and Lu,  path relinking (PR) algorithm from \cite{lai14} and learning hybrid-based search (LHS) from \cite{jin14}. Like in the case of BCP, all these results are also extracted from \cite{jin14}. Last three columns contain best results, the
execution times for the proposed VNS and the difference between best VNS and previous best-known results.

\begin{table}[htbp]
  \centering
  \caption{\small{Comparison of the proposed VNS algorithm to other reference works on BMCP instances}}
  \tiny
\begin{tabular}{rrrrrrrrrrrrrrl}\hline
&&\multicolumn{2}{c}{\cite{pre08} }&\multicolumn{2}{c}{\cite{mal08} }&\multicolumn{2}{c}{\cite{lai13} }
&\multicolumn{2}{c}{\cite{lai14} }&\multicolumn{2}{c}{\cite{jin14} }&\multicolumn{2}{c}{VNS}&\\

Instance & best & $k$ & time & $k$ & time  & $k$ & time   & $k$ & time   & $k$ & time    & $k^*$          & time{[}s{]} & diff \\\hline
GEOM20   & 149  & 149 & 4    & 149 & 18    & 149 & 2      & 149 & 1      & 149 & 1.8     & 149            & 54.2701     & 0          \\
GEOM20a  & 169  & 170 & 2    & 169 & 9     & 169 & 15     & 169 & 7      & 169 & 0.5     & 169            & 3289.39977  & 0          \\
GEOM20b  & 44   & 44  & 0    & 44  & 5     & 44  & 0      & 44  & 0      & 44  & 0       & 44             & 0.02153     & 0          \\
GEOM30   & 160  & 160 & 0    & 160 & 1     & 160 & 0      & 160 & 0      & 160 & 0.1     & 160            & 5.7033      & 0          \\
GEOM30a  & 209  & 214 & 11   & 210 & 954   & 209 & 10     & 209 & 26     & 209 & 16.2    & 209            & 4123.7125   & 0          \\
GEOM30b  & 77   & 77  & 0    & 77  & 0     & 77  & 0      & 77  & 0      & 77  & 0       & 77             & 1.2923      & 0          \\
GEOM40   & 167  & 167 & 1    & 167 & 20    & 167 & 0      & 167 & 1      & 167 & 0.2     & 167            & 2107.13567  & 0          \\
GEOM40a  & 213  & 217 & 299  & 214 & 393   & 213 & 328    & 213 & 133    & 213 & 9       & 213            & 13192.2284  & 0          \\
GEOM40b  & 74   & 74  & 4    & 74  & 1     & 74  & 2      & 74  & 4      & 74  & 1.5     & 74             & 1821.8568   & 0          \\
GEOM50   & 224  & 224 & 1    & 224 & 1197  & 224 & 8      & 224 & 2      & 224 & 0.3     & 224            & 1671.3315   & 0          \\
GEOM50a  & 311  & 323 & 51   & 316 & 4675  & 314 & 40373  & 312 & 270860 & 311 & 1452.6  & 311            & 21685.93933 & 0          \\
GEOM50b  & 83   & 86  & 1    & 83  & 197   & 83  & 1200   & 83  & 723    & 83  & 72.1    & 83             & 14260.65607 & 0          \\
GEOM60   & 258  & 258 & 77   & 258 & 139   & 258 & 19     & 258 & 23     & 258 & 1.3     & 258            & 5780.4842   & 0          \\
GEOM60a  & 353  & 373 & 10   & 357 & 8706  & 356 & 38570  & 354 & 34580  & 353 & 9007.1  & 353            & 23989.30317 & 0          \\
GEOM60b  & 113  & 116 & 12   & 115 & 460   & 113 & 104711 & 113 & 63579  & 113 & 910.7   & 114            & 16381.0242  & 1          \\
GEOM70   & 266  & 277 & 641  & 272 & 1413  & 270 & 7602   & 266 & 130844 & 266 & 2534    & 267            & 12384.2772  & 1          \\
GEOM70a  & 465  & 482 & 315  & 473 & 988   & 467 & 38759  & 466 & 6952   & 465 & 36604.9 & \textbf{463} & 20686.36647 & -2         \\
GEOM70b  & 115  & 119 & 55   & 117 & 897   & 116 & 213545 & 116 & 26110  & 115 & 3640.7  & 116            & 16783.2271  & 1          \\
GEOM80   & 379  & 398 & 361  & 388 & 132   & 381 & 212213 & 380 & 34493  & 379 & 357.8   & 379            & 16538.80857 & 0          \\
GEOM80a  & 357  & 380 & 109  & 363 & 8583  & 361 & 41235  & 358 & 41772  & 357 & 43403   & \textbf{355} & 29208.93163 & -2         \\
GEOM80b  & 138  & 141 & 37   & 141 & 1856  & 139 & 255    & 138 & 705    & 138 & 46.5    & 138            & 13704.121   & 0          \\
GEOM90   & 328  & 339 & 44   & 332 & 4160  & 330 & 4022   & 328 & 134941 & 328 & 162.2   & 329            & 18760.85243 & 1          \\
GEOM90a  & 372  & 382 & 13   & 382 & 5334  & 375 & 10427  & 372 & 282456 & 372 & 16782.1 & 373            & 24087.21087 & 1          \\
GEOM90b  & 142  & 147 & 303  & 144 & 1750  & 144 & 211366 & 144 & 14648  & 142 & 7680.8  & 142            & 19996.89127 & 0          \\
GEOM100  & 404  & 424 & 7    & 410 & 3283  & 404 & 40121  & 404 & 16355  & 404 & 64.9    & 404            & 14816.92453 & 0          \\
GEOM100a & 429  & 461 & 26   & 444 & 12526 & 442 & 381    & 436 & 9108   & 429 & 78363.1 & 429            & 35663.42977 & 0          \\
GEOM100b & 153  & 159 & 367  & 156 & 3699  & 156 & 213949 & 156 & 86308  & 153 & 10840.1 & 156            & 24776.3847  & 3          \\
GEOM110  & 375  & 392 & 43   & 383 & 2344  & 381 & 183    & 375 & 25401  & 375 & 1598.8  & 375            & 22997.44417 & 0          \\
GEOM110a & 478  & 500 & 29   & 490 & 2318  & 488 & 926    & 482 & 9819   & 478 & 49457.1 & 480            & 46954.99173 & 2          \\
GEOM110b & 201  & 208 & 5    & 206 & 480   & 204 & 944    & 201 & 47653  & 201 & 5388.4  & 202            & 22076.94307 & 1          \\
GEOM120  & 396  & 417 & 9    & 396 & 2867  & -   & -      & 396 & 15341  & 396 & 626.1   & 396            & 17919.32823 & 0          \\
GEOM120a & 536  & 565 & 41   & 559 & 3873  & 554 & 1018   & 539 & 45147  & 536 & 69518.6 & 539            & 59717.3869  & 3          \\
GEOM120b & 187  & 196 & 3    & 191 & 3292  & 189 & 213989 & 189 & 14371  & 187 & 8025.8  & 190            & 22835.9246  & 3

\end{tabular}

\label{table:BMCP2}

\end{table} 
 From the Table \ref{table:BMCP2} it can be seen that FCNS, EA, MITS and PR could achieve most of the best known solutions for small instances up to 50 vertices, and LHS and VNS achieve all of the best known solutions for small instances. MITS and PR fails to achieve best known solution only for one small instance (GEOM50a).

 For the middle-size instances (60 to 90 vertices), the situation is slightly different; FCNS and EA could achieve only one best known solution (instance GEOM60), while MITS could achieve two best known solutions for two instances with 60 vertices. For other middle-size instances, MITS achieves nearly best known solutions. For this class of instances, PR could achieve 6 best known solutions and in other cases is close to the best ones. LHS achieves all best known solutions except in ten cases and in two cases close to the best ones. VNS achieves five best known solutions, in five cases VNS achieve solutions close to the best ones and in two cases VNS finds new best solutions.

For the large-size instances, PR could not find any best known solution, while EA and MITS achieve only one best known solution. PR achieves four best known solutions and for other large-size instances achieve solutions relatively close to the best ones. While LHS achieves all best knows solutions for large instances, VNS achieve best ones for six instances and for the rest nearly best solutions.

 Regarding execution times, it is obvious that they are quite different, since different approaches provide results with different $k$ values. Additionally, some of the algorithms (like EA and VNS) spend additional time for the construction phase, while other algorithms (MITS, PR and LHS) start  with the previous best known solutions.

 \subsection{Justification of the usage of the criteria in the VND procedure}\label{section:criteriaexperimental}
 In order to further analyse the behaviour of the proposed VND procedure and to justify the usage of the criteria for ordering vertices, additional experiments are performed with three large BMCP instances (namely GEOM110, GEOM110a and GEOM110b). Recall that in the VND procedure three criteria are combined for determining the ordering: the first criterium is the total number of conflicts for the vertex, the second  is the distance of the color assigned to the vertex from the middle and the third one is the geometric mean of the total sum of the weights of the edges incident with the vertex and the maximal edge distance for that vertex.
 In the experiment, total of eight possible different variants of the overall criterium are analysed: each of the mentioned three criteria  is or is not used.

 Table \ref{table:criteria} provides the experimental results obtained in all of these eight cases. Each case is denoted as $XYZ$, where $X,Y$ and $Z$ belongs to $\{0,1\}$, indicating if the criterium is used (value 1) or not used (value 0). For each case, best found and average results are shown. The time limit for this experiment is set to 3000s.
\begin{table}[htbp]
  \centering
  \caption{\small{Comparison of the various combinations of the criteria used in the VND}}
  \label{table:criteria}\tiny
  \begin{tabular}{|l|l|l|l|l|l|l|l|l|}\hline
   & \multicolumn{2}{|c|}{000}               &\multicolumn{2}{|c|} {001}               & \multicolumn{2}{|c|}{010}                & \multicolumn{2}{|c|}{011}                \\\hline
   instance&best&avg&best&avg&best&avg&best&avg\\\hline
Geom110  & 376 & 379.16667 & 375 & 378.53333 & 375 & 378.96667 & 375 & 378.46667 \\
GEOM110a & 486 & 491.23333 & 487 & 491.23333 & 483 & 487.86667 & 483 & 488.33333 \\
GEOM110b & 201 & 204.06667 & 201 & 204.3     & 202 & 204.1     & 202 & 203.96667\\\hline
   & \multicolumn{2}{|c|}{100}                &    \multicolumn{2}{|c|}{101}      &   \multicolumn{2}{|c|}{110}
      &    \multicolumn{2}{|c|}{111}               \\\hline
    instance     &best&avg&best&avg&best&avg&best&avg\\\hline
Geom110  & 376 & 378.2     & 376 & 378.03333 & 376 & 378.56667 & 375 & 377.9    \\
GEOM110a & 482 & 489.83333 & 484 & 488.96667 & 483 & 487.73333 & 481 & 486.8333 \\
GEOM110b & 203 & 204.6     & 202 & 204.6     & 202 & 204.16667 & 202 & 204.0333\\\hline
\end{tabular}

  \end{table} 
  From the table \ref{table:criteria}, it can be seen that the variant proposed in this paper (column 111 - all three criteria are used) tends to be the best one, although some other variants also gives good results. For the first  instance (GEOM110), in the variants  denoted as 001, 010 and 011 the best result is also  achieved, but the average result is worse than in the proposed variant (111).  For the second instance (GEOM110a), in the proposed variant (111) both the best known and the average values are better than in the other cases. For the last instance (GEOM110b), the first two variants achieves better best result than the proposed one, but the average value obtained by the proposed variant is better.

  These results indicate that the usage of the proposed combination of the criteria is justified, having in mind that some other variants can also reach high quality results.

\section{Conclusions}

In this paper we present the VNS algorithm for solving two generalizations of the vertex coloring problems: bandwidth coloring problem and
 multiple bandwidth coloring problem.
 Since BCP and MBCP enjoy many  applications, presented algorithm and achieved results are of a great interest for both theory and practice.

 In the shaking procedure,  an increasing number of vertices are permuted, forming the new solution which is subject of the further improvement in  the
 VND. In the VND procedure, vertices are arranged in a way to increase the probability of successful recoloring.  This approach of the arrangement of the
 vertices splits the local search in a series of disjoint procedures, enabling better choices of the vertices which are addressed to re-color. The overall
 criterion for the ordering the vertices is based on  the number of conflicts of each vertex, combined with two additional criteria.

The algorithm is tested on the common used instances. In the case of BCP, VNS achieves many previous best-known results and in the case of BMCP, VNS obtains two new best solutions in a reasonable time.  Experimental results indicate that the proposed VNS becomes one of state-of-the-art methods for
solving BCP and BMCP.

 The further investigation of this problem can include parallelization and running on some powerful multiprocessor system, as well as the application of
 the proposed method for solving some similar real-life problems.

\bibliographystyle{elsarticle-num}




\end{document}